\documentstyle[twocolumn] {article}
\title
{ON THE POSSIBILITY OF QUANTUM-MECHANICAL INTERPRETATION THE
 RELATIVISTIC EFFECT OF ENERGY INCREASE IN A PARTICLE FREELY MOVING IN
 VACUUM}

\author{G. N. Gestrina\\
Usikov Institute of Radio Physics and Electronics\\
of
National Academy of Ukraine                     \\
12, Proskura st., Kharkov, 61085, Ukraine       \\
E-mail: ire@ire.kharkov.ua}

\date{}
\begin{document}
\maketitle

\begin{abstract}
The relativistic effect of energy increase in a particle freely
moving in vacuum is discussed on the basis of quantum field theory
and probability theory using some ideas of super-symmetrical
theories. The particle is assumed to consist of a "seed" whose
energy is equal to the  particle rest energy and whose pulse is
equal  to the product of the particle mass by its velocity and of
a "fur coat" - the system of virtual quanta of the material field
- vacuum. Each of these quanta possesses the same energy and pulse
as the "seed" but have no mass. The system of the quanta is in a
state being the superposition of quantum states with energies and
pulses multiple of the "seed" energy and pulse. The virtual quanta
is created (or destroyed) in of such states. The probability of
creating a quanta in any state is the inverse of the relativistic
factor, and the average number of the quanta making up the "fur
coat" with a "seed" is equal to this particular factor. The
kinetic energy and the relativistic addition to the particle pulse
are interpreted as the average magnitude of the energy and the
pulse in the system of the virtual quanta that constitute the
particle "fur coat".
\end{abstract}

The Einstein formula linking the energy and the mass of a particle
freely moving in  vacuum, certainly, is the central formula of
special
      theory of  relativity [1]. It has played a tremendous role for a
       deeper understanding  of the material world surrounding us and has
        had an enormous practical  application. This formula shows that
     the energy of a relativistic particle  increases with its velocity
      whereas its mass remains constant [2].

On the modern stage of physical science it becomes possible to describe
 this  effect by means the quantum field theory. In this case some ideas of
   super-symmetrical theories turned out to be rather useful [3].

1. First it will be shown how by means of a simple mathematical
 transformation the Einstein formula may be expressed in the form allowing
  its interpretation on the quantum-mechanical basis.

The kinetic energy of a relativistic particle with mass $m$ moving with
  constant velocity $\bf v$ may be presented as follows [4]:
    \begin{eqnarray}
E_{kin}=mc^{2}\left(
\frac{1}{\sqrt{1-\left( \frac{v}{c} \right)^2}}-1 \right)=  \nonumber \\
=mc^2\sqrt{1-\left( \frac{v}{c} \right)^2}
\frac{1-\sqrt{1-\left( \frac{v}{c} \right)^2}}{1-\left(
\frac{v}{c}\right)^2}=\\
=mc^2\sqrt{1-\left( \frac{v}{c} \right)^2}\sum_{n=1}^{\infty}
n\left[ 1-\sqrt{1-\left( \frac{v}{c} \right)^2}\right]^n, \nonumber
\footnote{Allowing $1-\sqrt{1-(v/c)^2}=u$, we notice that\\
$\left[1-\sqrt{1-(v/c)^2}\right] /
\left[1-(v/c)^2\right]=u/(1-u)^2=\sum_{n=1}^\infty nu^n$.}
     \end{eqnarray} where
$c$ is velocity of light in vacuum.  Supposing further that
\begin{equation} E_n=nmc^2 \; \; \;   (n=1,2,\ldots )
    \end{equation}
 and
    \begin{eqnarray}
P_n=P_0(1-P_0)^n \; \; \; \\
 (n=0,1,2,\ldots,\; 0<P_n<1,\; \sum_{n=0}^\infty P_n=1), \nonumber
    \end{eqnarray}

 where
    \begin{equation}
P_0=\sqrt{1-\left( \frac{v}{c} \right)^2},
    \end{equation}
 we get
    \begin{equation}
    E_{kin}=\sum_{n=1}^\infty E_nP_n,
    \end{equation}
The full energy $E$ of a relativistic particle can be
written as:
    \begin{equation}
    E=mc^2+\sum_{n=1}^\infty E_nP_n,
    \end{equation}
 and its pulse $\bf p$  due to the relativistic relationship
${\bf p}=E{\bf v}/c^2$ will take the form:
    \begin{equation}
    {\bf p}=m{\bf v}+\sum_{n=1}^\infty {\bf p}_nP_n,
    \end{equation}
where
    \begin{equation}
    {\bf p}_n=nm{\bf v}\; \; \; (n=1,2,\ldots).
    \end{equation}

    As it follows from relations (6) and (7) a relativistic particle with
 mass $m$ freely moving in vacuum at constant velocity $\bf v$,
 can be considered as an aggregate consisting of a "seed" particle with
 energy $E_0=mc^2$  and pulse ${\bf p_0}=m\bf v$
 and an infinite system of virtual quanta of the material field, i.e.,
 the vacuum ("fur coat"), each of them also having energy
 $E_0=mc^2$ and pulse ${\bf p}_0=m\bf v$. This
 system of quanta is in a state representing the superposition of
 $n^{thes}$ quantum states, characterized by
 energies $E_n=nmc^2$ and pulses ${\bf p}_n=nm{\bf v}\; \; \;
 (n=1,2,\ldots)$.

   The interpretation of $P_n$
 values as some probabilities will be given below.  The accumulation by
 the "seed" particle of such a "fur coat" consisting of virtual quanta is
    very much alike the crystallization process in a saturated solution
 of a substance after an isolated crystal "seed" of the same substance
 has been introduced into it. In both cases a similar reaction of the
 ambient medium (a vacuum or a saturated solution) is observed with the
 only difference that in the first case virtual field quanta possessing
 the same energy and pulses have no mass whereas crystals obtained from
 the saturated  solution represent precise copies of "seed", i.e., of
 an isolated crystal introduction from outside.

Consequently, already within the framework of the special relativity
theory vacuum acquires the sense of a material field. The virtual
quanta of this field are bosons and  $n^{thes}$  quantum states of
the virtual quantum quantum system are $n^{thes}$    boson states.

 2. Let us correlate each $n^{th}$ boson state of this system
    with the $n^{th}$ excited state of a harmonic oscillator with frequency
 $\omega=E_0/\hbar$ ($\hbar=h/2\pi$, $h$--Plank constant).
   However in the situation described the lowest energy
 state of the material field for a particle (i.e., the vacuum) has a
 zero energy whereas the lowest energy value of the harmonic oscillator
 is $\hbar\omega/2$. To obtain a correspondence
 between the bosonic states of virtual quanta and the excited state of
 the harmonic oscillator through rejecting $\hbar\omega/2$ energy
 is impossible to neglect zero oscillations of the vacuum since they
 are wholly real [3].

    In super-symmetrical theories where bosons and fermions are considered
 together the infinite positive energy of zero boson oscillations is fully
 compensated by the infinite negative energy of fermion oscillations
therefore the energy of a single boson-fermion vacuum is zero. What is
 considered in super-symmetrical theories is a super-symmetrical harmonic
 oscillator. Hamiltonian of such an oscillator is the sum of
 Hamiltonians
 of two noninteracting harmonic oscillators corresponding to boson and
 fermion degrees of freedom. Frequencies of these oscillators must be equal.
 The lowest energetic state of the super-symmetrical oscillator is zero [3].

    In some situations when fermion degrees of freedom are absent
 there exists, however, the possibility to use the ideas concerning the
 zero energy of the vacuum and two harmonic oscillators with equal
 frequencies. Thus, in the described situation it is  sufficient to
 suppose that in the $n^{th}$ boson state there are $n_+=N_n+n$
  bosons with $E_0>0$ and $n_-=N_n$
  bosons with $E_0<0$ $(E_0=\pm mc^2)$. The difference $n$
  $(n=n_+-n_-)$
  determines the energy of the $n^{th}$  boson state. $N_n$ virtual
  quanta with $E_0>0$ compensate  $N_n$ ones with $E_0<0$. $N_n$   is a
 random positive integer therefore all the $n^{ths}$ quantum boson
 states are  degenerate.

     Let the basis vector of the $n^{ths}$
quantum states of the virtual quantum system be represented in the
form [3]:
    \begin{equation}
    |n_+,n_-\rangle=|\psi_n\rangle.
    \end{equation}
    Here $n_+$ and $n_-$ are
 population numbers for quanta with $E_0>0$ and $E_0<0$,
 respectively, $|\psi_n\rangle $ is the vector
of the $n^{th}$ quantum state. $|N_n,N_n\rangle $
 vector will describe the vacuum field state. It is quite
 clear that the energy of the vacuum is zero.

     Now it is necessary to
 consider two harmonic oscillators with the same frequency
 $\omega=|E_0|/\hbar$. Let $q_1,p'_1$, $m_1$
  and $H_+$ be the generalized coordinate,
 the  pulse, the mass and the Hamiltonian of the first oscillator
    corresponding to boson degrees of freedom with $E_0>0$,
     and $q_2,p'_2, m_2$ and $H_-$
  be the generalized coordinate, the  pulse, the mass
 and the Hamiltonian of the second oscillator corresponding to boson
 degrees of freedom with $E_0<0$, that is
    \begin{equation}
    H_+=\frac{1}{2}\left(
    \frac{p_1^{'\,2}}{m_1^2}+m_1\omega q_1^2
    \right), \,\, m_1=+\frac{|E_0|}{c^2}
    \end{equation}
and
    \begin{equation}
    H_-=\frac{1}{2}\left(
       \frac{p_2^{'\,2}}{m_2^2}+m_2\omega q_2^2 \right),
      \,\, m_2=-\frac{|E_0|}{c^2} .
       \end{equation}

 Correspondingly the
energy levels of these oscillators have the form
    \begin{equation}
    E_{n_+}=\hbar \omega (n_++\frac{1}{2}) \; \; \; (n_+=n,n+1,\ldots)
    \end{equation}
 and
       \begin{equation}
    E_{n_-}=\hbar \omega (n_-+\frac{1}{2}) \; \; \; (n_-=0,1,\ldots)
    \end{equation}

 Creation and  destruction operators for bosons with $E_0>0 \;
 \;(C_+^{\pm})$  and for bosons with $E_0<0 \; \;(C_-^{\pm})$
 are defined as follows [5]:
    \begin{equation} \label{f14}
    C_+^{\pm}=(2\hbar \omega m_1)^{-1/2}(p'_1 \pm im_1 \omega q_1)
    \end{equation}
 and
    \begin{equation} \label{f15}
    C_-^{\pm}=(-2\hbar \omega m_2)^{-1/2}(p'_2 \pm im_2 \omega
        q_2).
    \end{equation}
They satisfy the ordinary commutation relations
    \begin{equation} \label{f16}
    [C_{\pm}^-,C_{\pm}^+]=1, \,\,\,\,  [C_-^{\pm},C_+^{\pm}]=0.
    \end{equation}

Their effect on state vectors $|n_+,n_-\rangle$
    will be the following:
    \begin{equation} \label{f17}
    C_+^+|n_+,n_-\rangle =\sqrt{n_++1}\;|n_++1,n_-\rangle;
    \end{equation}
    \begin{equation} \label{f18}
    C_+^-|n_+,n_-\rangle =\sqrt{n_+}\;|n_+-1,n_-\rangle;
    \end{equation}
    \begin{equation} \label{f19}
    C_-^+|n_+,n_-\rangle =\sqrt{n_+1}\;|n_+,n_-+1\rangle;
    \end{equation}
    \begin{equation} \label{f20}
    C_-^-|n_+,n_-\rangle = \sqrt{n_-} \; |n_+,n_--1 \rangle .
    \end{equation}

Thus, the creation operator $C_+^+$ acting on
$|n_+,n_-\rangle=|\psi_n\rangle$  vector transforms it to
$|n_++1,n_-\rangle=|\psi_{n+1}\rangle$ vector, whereas the
annihilation operator  $C_+^-$ transforms the same vector to
$|n_+-1,n_-\rangle=|\psi_{n-1}\rangle$ vector.

Similarly the creation operator $C_-^+$
acting on $|n_+,n_-\rangle=|\psi_n\rangle$ vector transforms it into
$|n_+,n_-+1\rangle=|\psi_{n-1}\rangle$ vector, at
 the same time the annihilation $C_-^-$ operator
 transforms the same vector into
 $|n_+,n_--1\rangle=|\psi_{n+1}\rangle$ vector.

 It  can be seen that $|n_++1,n_-\rangle$ and $|n_+,n_- -1\rangle$
 vectors describe equivalent energy states characterized by
 $|\psi_{n+1}\rangle$ vector, and $|n_+-1,n_-\rangle$
 and $|n_+,n_-+1\rangle$ vectors describe those
 characterized by  $|\psi_{n-1}\rangle$ vector.

 Operator
    \begin{equation} \label{f21}
    C_+^+C_+^-= \hat{n}_+
    \end{equation}
 will be an operator of the boson number with $E_0>0$,
  whereas operator
    \begin{equation} \label{f22}
    C_-^+C_-^-= \hat{n}_-
    \end{equation}
will define the number of bosons with $E_0<0$.

 Correspondingly $H_{\pm}$ Hamiltonians will be
 expressed through $\hat{n}_{\pm}$ operators as
follows:
    \begin{equation} \label{f23}
    H_+=\hbar \omega (\hat{n}_++\frac{1}{2})
    \end{equation}
and
    \begin{equation} \label{f24}
    H_-=-\hbar \omega (\hat{n}_-+\frac{1}{2}).
    \end{equation}

Since there is no interaction between the oscillators their common
Hamiltonian $H$ will be equal to the sum of $H_+$ and $H_-$
Hamiltonians, i.e.,
    \begin{equation} \label{f25}
    H=\hbar \omega (\hat{n}_+-\hat{n}_-).
    \end{equation}

 The oscillator corresponding to this Hamiltonian
 is not super-symmetrical
    (in a super-symmetrical harmonic oscillator all the states except
 zero one are doubly degenerate [3]).

 It is obvious that
    \begin{equation} \label{f26}
    H\;|\psi_n\rangle =E_n\; |\psi_n\rangle  \;\;\;(n=1,2,\ldots).
    \end{equation}

 This means that each of $|\psi_n\rangle $
 vectors is an eigenvector of $H$ Hamiltonian
 pertaining to the eigenvalue of $E_n=nE_0$  energy.

 The $|\psi_n\rangle $ vectors form an orthonormal system of
 vectors:
    \begin{equation} \label{f27}
    \langle \psi_n|\psi_{n'}\rangle =\delta_{nn'}
    \end{equation}
($\delta_{nn'}$  is the
 Kroneker symbol). Any admissible state $|\psi \rangle$
vector of virtual quantum system is a superposition of
$|\psi_n \rangle$   vectors, i.e.,
    \begin{equation} \label{f28}
    |\psi \rangle =\sum_n \alpha_n \;|\psi_n \rangle ,
    \end{equation}
where $\alpha_n$  factors are determined
 from relations
    \begin{equation} \label{f29}
    \alpha_n=\langle \psi | \psi_n \rangle
    \end{equation}
and satisfy the normalizing condition:
    \begin{equation} \label{f30}
    \sum_n |\alpha_n|^2=1.
    \end{equation}
The average value of virtual
 quantum energy in the state described by $|\psi \rangle$
vector equal:
    \begin{equation} \label{f31}
    \bar E =\langle \psi | H | \psi \rangle =\sum_n
    |\alpha_n|^2E_n.
     \end{equation}

    3. Let us clear up the meaning of $P_n$-values
 and their connection with $\alpha_n$. For this
 purpose a complete system of independent events $A_n$  and their opposite
 events $\bar A_n$ will be introduced (Bernoulli's scheme)[6]. $A_n$-event will
 mean the creation of a virtual quantum with  $E_0>0$
  (or the destruction of a virtual quantum with $E_0<0$)
 in the $n^{th}$ quantum state with $P_0$
 probability. At the same time let us assume that  $\bar A_n$
means the destruction of a virtual quantum with $E_0>0$
(or the creation of a virtual quantum with $E_0<0$)
in the $n^{th}$ quantum state with a probability of $1-P_0$.

  $P_n=P_0(1-P_n)^n$ values will mean the
 probabilities of that the creation of the virtual quantum with $E_0>0$
has not taken place in quantum states from $0$ to $n-1$ but
    only in the $n^{th}$ quantum state (or the probability that the
 quantum with $E_0<0$ was not destructed in quantum
 states from $0$ to $(n-1)^{th}$ but was destructed only in the $n^{th}$
 quantum state).

Let us determine the value
    \begin{equation} \label{f32}
    \bar n+1=\sum_{n=1}^{\infty}nP_n+1=
    P_0 \sum_{n=1}^{\infty}n(1-P_0)^{n-1}=1/P_0.
    \end{equation}

 If $\bar n+1$ is multiplied by $mc^2$
one gets the complete energy $E$ of a relativistic particle, whereas
the production of the multiplying $\bar n+1$
  by $m\bf v$ gives its pulse. Multiplying $\bar n$
  by $mc^2$ we get an expression  for $E_{kin}$
of the relativistic particle. Consequently  $\bar n$
  value can be interpreted as an average number
 of virtual quanta forming "fur coat" of the relativistic particle whereas
$E_{kin}$ is an average value of energy of these quanta.
    Comparing the expression for $E_{kin}$ (5) with expression
 (31) for the  average value of energy of virtual quanta in a state that
 is a superposition of the $n^{ths}$
quantum states described by $|\psi \rangle$ vector one can conclude that
    \begin{equation} \label{f33}
    |\alpha_n|=\sqrt{P_n}.
    \end{equation}

    4. Let us write the expression for the de'Broglie wave of the
 relativistic particle as follows:
    \begin{eqnarray} \label{f34}
    \psi = C\exp[-i(Et-{\bf pr}/\hbar)]=\\
    C\exp\{-i[(E_0t-{\bf p}_0{\bf r})+
    \sum_{n=1}^{\infty}(E_nt-{\bf p}_n{\bf r})P_n]/\hbar
    \}=\nonumber \\
    =\psi_0 \prod_{n=1}^{\infty} \psi_n ,  \nonumber
    \end{eqnarray}
where
    \begin{equation} \label{f35}
    \psi_0=C_0\exp[-i(E_0t-{\bf p}_0{\bf r})/\hbar]
    \end{equation}
is the de'Broglie wave of the "seed" particle and
    \begin{eqnarray} \label{f36}
    \psi_n=C_n\exp[-i(E_nt-{\bf p}_n{\bf r})/\hbar]=  \\
    =C_n\exp[-i(E_0t-{\bf p}_0{\bf r})nP_n/\hbar]  \nonumber
    \end{eqnarray}
are the de'Broglie waves of virtual quanta,
    \begin{equation} \label{f37}
    C=C_0\prod_{n=1}^{\infty} C_n,
    \end{equation}
$\bf r$ is a radius-vector of any point of the space.

Expression (34) shows that there is no interaction between the
 "seed" and the virtual quanta  as well as  between the latters themselves.

 All the de'Broglie waves $\psi_0$  and
$\psi_n\;\;(n=1,2,\ldots)$ have the same phase velocity $v_p=c^2/v,$
the frequencies $\omega_0=E_0/\hbar$ are $\omega_n=E_nP_n/\hbar=
n\omega_0P_n \;\; (n=1,2,\ldots)$ and the lengths
$\lambda_0=\hbar/p_0$ and $\lambda_n= \hbar/p_nP_n \;\;
(n=1,2,\ldots)$, where $\omega_0$ and $\lambda_0$ are the frequency
and the  de'Broglie wavelength of the
 "seed" particle whereas $p_0$ is its pulse.

    But neither the "seed" particle nor the virtual quanta can be
 separately fixed by a device.  The latter fixes the  relativistic
 particle as a whole.  This follows from the fact that diameters of
 the   diffraction coils on the electronogram are expressed through the
 relativistic value for the de'Broglie wavelength
 $\lambda=\hbar \sqrt{1-(\frac{v}{c})^2}/mv$
 of a particle.

 Thus the quantum field theory combined with some ideas
 of modern super-symmetrical theories allows to consider "fur coat" of a
 relativistic particle freely moving in vacuum as a virtual quantum
 collection of the vacuum each of which has the same energy and pulse
 as the "seed" particle.

 The boson degrees of freedom discussed above
    are probably those internal degrees of freedom which P. Dirak [7,
 8] had in mind suggesting the availability of the dynamic structure in
 the relativistic particle.

 From the standpoint of the probability
    theory the "fur coats" formation of the relativistic particle out of
 vacuum virtual quantum is regarded as a random event.

    1. A. Einstein, Ann. d. Phys. Bd. 17, 851 (1905).

    2. L.B. Okun', UFN, 158, 511 (1989).

    3. L.E. Gendenshtein, I.V. Krive, UFN, 146, 553 (1985)

    4. G.N. Gestrina, Reports of Uk AS, ser "A", 9, 55 (1990)

    5. V.B.Berestetsky, E.M.Lifshits, L.P.Pitaevsky. Quantum
    Electrodynamics.  Moscow; Nauka, 1974 (in Russian)

    6. E.S.Ventsel. Theory Probabilities.  Moscow; Nauka, 1984 (in
    Russian)

    7. P.A.M. Dirac, Proc.Roy.Soc., A,322,435 (1971)

    8. P.A.M. Dirac, Proc.Roy.Soc., A,328,1 (1972)

\end{document}